# Fabrication of Triangular Nanobeam Waveguide Networks in Bulk Diamond Using Single-Crystal Silicon Hard Masks

I. Bayn[1§], S. Mouradian[1§], L. Li[1], J. A. Goldstein[1], T. Schröder[1], J. Zheng[1], E. H. Chen[1], O. Gaathon[1], M. Lu[2], A. Stein[2], C. A. Ruggiero[2], J. Salzman[3], R. Kalish[4], and Dirk Englund[1]

*1. Department of Electrical Engineering and Computer Science, and Research Lab of Electronics, Massachusetts Institute of Technology, 77 Massachusetts Ave., Building 36. Cambridge, MA 02139, USA*

*2. Center for Functional Nanomaterials, Brookhaven National Laboratory, Upton, NY 11973, USA*

*3 Department of Electrical Engineering and Microelectronics Research Center, Technion Haifa, 32000, Israel*

*4 Department of Physics and Solid State Institute, Technion Haifa, 32000, Israel*

A scalable approach for integrated photonic networks in single-crystal diamond using triangular etching of bulk samples is presented. We describe designs of high quality factor ($Q=2.51\times10^6$) photonic crystal cavities with low mode volume ($V_m=1.062\times(\lambda/n)^3$), which are connected via waveguides supported by suspension structures with predicted transmission loss of only 0.05 dB. We demonstrate the fabrication of these structures using transferred single-crystal silicon hard masks and angular dry etching, yielding photonic crystal cavities in the visible spectrum with measured quality factors in excess of $Q=3\times10^3$.

§ These authors contributed equally.

The negatively charged nitrogen-vacancy color center in diamond (NV) has attracted wide interest as a spin-based quantum memory that can be initialized, manipulated, stored, and measured optically[1]. Spatially separated NVs can interact via photons[2], and entanglement between two proximal NVs was recently demonstrated[3]. Coupling NVs to dielectric nanocavities can significantly improve the efficiency of coherent transitions via the NV zero phonon line (ZPL). To this end, cavities have been fabricated using hybrid material approaches[4], as well as using poly-crystal[5] and single-crystal diamond[6,7]. Recently, to utilize the superb NV quantum properties of bulk single-crystal diamond[8], some of us proposed triangular nanobeam designs and prototyped free-standing devices with Focused Ion Beam milling (FIB)[9]. Subsequently, such beams were also realized by Burek *et al* (Ref. 10) using an angular dry etching approach[11] which enables photonic crystal nanocavities. Here, we present an expanded design to produce spatially scalable networks of cavities connected via triangular nanobeams[7] suspended by diamond waveguide support structures. We experimentally demonstrate such suspended cavity-waveguide structures using a novel fabrication process that employs a high-quality mono-crystalline silicon mask with excellent spatial resolution of apertures down to 25 nm, position precision of less than 2 nm, and high (over 99:1) selectivity in diamond etching. Such nanocavities realized in diamond would also enhance other color centers, such as the silicon-vacancy recently shown to have narrower ZPL with significantly less spectral diffusion than the NV.[7,11]

As shown in Fig. 1a, the diamond waveguide - oriented in the z-axis - has a triangular profile in the **xy** plane with height $H$ and width $W$. A lattice of rectangular holes with cross sections $W_x \times W_z$ in the z-direction produces a 1D photonic crystal (PC). As shown previously[9], introducing a defect in this periodic structure forms a cavity. Here, we consider a defect region in which the lattice constant increases parabolically from $a_0$ in the cavity center to $a$ in the waveguide regions, according to $a_N=a_0+(N/N_0)^2 \times k \times a$ for $|N|\leq N_0=7$, as shown in Fig. 1b.[12] We used 3D Finite-Difference-Time-Domain (3D-FDTD) simulations to optimize cavities for high quality factor ($Q$) and low mode volume ($V_m$). Under this parabolic model, we found an optimal $Q/V$ ratio for $a_0=0.9a$ and $k=0.1$. The cavity $Q$ is limited by the upward ($+\hat{y}$) photon losses (our simulations show that the upward intensity loss is 5-20 times larger than downward loss). The design was optimized by a gradient search with sweeps over parameters $W_x$, $W$, and $H$, which predominantly affect $Q$ and $V_m$ (see Fig. 1c). For the optimal mode parameters found in this way ($H=1.35a$, $W=2a$, $W_x=0.525W$, $W_z=0.5a$), we obtain a mode at frequency $a/\lambda \approx 0.339$ with $Q=2.51 \times 10^6$, $V_m=1.062 \times (\lambda/n)^3$. For an NV center that is located at the fundamental mode maximum ($x=z=0$ and $y=0.14a$, see Fig. 1d) and oriented with the electric field, this cavity would produce an enhancement of the ZPL transition rate by a factor $F_{cav}=(3/4\pi^2) \times (Q/V_m) \approx 1.79 \times 10^5$. Second, a linear heterostructure cavity in which $a_N=0.9a$ for $0\leq|N|\leq6$ and $a$ otherwise is also explored.[9] Previous works[9] report design with $Q$ of over $2.2 \times 10^4$, and here we show that a theoretical value of $Q > 1.1 \times 10^5$ can be obtained with geometry $H=a$, $W=2a$, $W_x=0.5W$, $W_z=0.5a$. Although the simulated $Q$ are smaller than that of the parabolic cavity, the linear heterostructure design is added for fabrication simplicity, as it requires less strict control of the grating pitch.

We fabricated the designed nanobeam structures in single-crystal diamond grown by chemical vapor deposition (CVD, Element6) with a nitrogen defect density of ~1 ppm and a surface roughness (Ra) of less than 30 nm. High-quality hard masks were fabricated on silicon-on-oxide (SOI) wafers with a silicon thickness of 270 nm. Masks were written into undiluted ZEP 520A e-beam resist - spun onto the SOI wafer at 4 Krpm – with a JEOL JBX-6300FS at 100 kV with 400 pA current at a dosage of 600 μC/cm$^2$. The resist was developed in -25°C Hexyl Acetate for 90 s. Following resist development, the silicon was dry-etched by Oxford Instruments Plasmalab 100 using a mixture of $SF_6$:$O_2$=40:18 at -100°C with ICP (RF) power of 800 W (15 W) for 30 s (see Fig. 2a1).

Post mask etch, patterned Si membranes were removed from the SOI substrate via an undercut in 49% hydrofluoric acid (HF) for 15 minutes. This undercut was facilitated by square etch holes as seen in Fig. 2a. The HF was slowly diluted with DI-water, and the masks were steered from the etched substrate with a directed water stream, and each was encapsulated in a separate droplet on a Teflon substrate. The droplets were then evaporated at 60°C leaving millimeter-scale mask sitting on Teflon (see Fig. 2a-3). Bringing the diamond sample into contact with the mask resulted in electrostatic attraction, transferring the mask from the Teflon to the diamond. Mask-diamond conformity was achieved through several cycles of mask refloat and drying with a methanol droplet released from a pipette on the diamond surface (see Fig.2a-4).

Post transfer, the diamond was etched using an inductively coupled oxygen plasma etcher (ICP) at flow rate of 14 sccm, working pressure of 3 mTorr with ICP and RF powers of 1000 W and 60 W, respectively, and a sample temperature of 60°C. The sample was first etched orthogonally to the surface with an etch rate of 0.375 μm per minute (see Fig.2a-4). The vertical etch was followed by an angular dry etch in a aluminum–made triangular Faraday cage[10] with rectangular base of 20 mm×12.7 mm and 6.4 mm high. The sample sitting in cage's center is exposed to the plasma through aluminum mesh with 750 μm pitch of 250 μm thick aluminum interwoven wires. To achieve more than 1 μm beam separation from the substrate (see Fig. 2a-5), the diamond was etched for 17 minutes as described above. Finally, the silicon mask was removed in KOH solution at 110°C for 1 h, leaving freestanding diamond beams (see Fig. 2a-6).

This Si hard-mask fabrication strategy is based on well-developed Si processing and so provides high resolution. Moreover, it ensures uniform coverage of the diamond surface while bypassing the challenges of fabricating on small diamond samples. Masks as large as 1.75 mm × 1.75 mm were transferred onto bulk diamond. Moreover, the diamond was only exposed to oxygen plasma during the dry etch step while avoiding other processing steps associated with electron beam lithography directly on diamond to minimize possible sources of diamond contamination that could degrade the spin and optical properties of NVs.

In Fig. 2b, a high magnification view of a broken beam is presented. The scanning-electron micrographs (SEMs) of angular (Fig. 2b) and vertical walls (Fig. 2a-4) show the low surface roughness obtained with the etching recipe presented above. The cross-sectional view of a beam prepared with focused ion beam (FIB) is shown in Fig. 2c. The fabricated angle $\theta$ (defined relative to horizontal) of *48°* and *51°* on each side with a bottom angle of *~28°* deviates from the expected triangular profile of *45°-90°-45°* dictated by the cage geometry, though we still see good agreement with the simulation results as discussed below. Initial examination shows that after angular over-etching of 42 minutes, triangular cross-sections can be achieved. Note that further modification of angular profile is possible by decreasing the angle of the cage grid; thus, decreasing $\theta$.

The asymmetry in the base angles of ~3° is produced by ±5° angular misalignment between the mask and the cage.

We produced cavities with the values of the lattice constant $a=220-240$ nm, and $W=2a$ for the parabolic modulation and linear heterostructures presented earlier (Fig. 1a). In both cases, the defect region was cladded with 22 periods of air holes on both sides. Fig. 3a shows a scanning electron micrograph (SEM) of a typical cavity. The shape of the grating holes was influenced by the aforementioned angular mask-cage misalignment. To examine the mechanical stability of the structure, we varied the air holes dimensions as follows $W_z=0.3-0.7a$. Only the holes with $W_z \leq 0.5a$ survived wet processing at high yield without significant beam bending.

We characterized these cavities by confocal spectroscopy, using an air objective with NA=0.9, 532 nm laser (0.2 mW of power in the laser focus spot and beam waist of $w_0$~250 nm) to excite native NV centers in the diamond whose fluorescence was used to pump the nanobeam's cavity modes. The excitation beam was circularly polarized, and fluorescence was collected through a polarizing beam splitter with a quarter waveplate used to remove polarization dependence of the collection. The fluorescence was spectrally filtered by 532 nm notch and 560 nm long pass filters, and spatially filtered by coupling into a single mode fiber. The signal was then detected by a silicon avalanche photodiode (APD, Perkin Elmer). Fig. 3b shows the spectrum of a characteristic parabolic cavity with $a=240$ nm and $W_{x,z}=(0.5W, 0.3a)$, indicating a quality factor of $Q$~3,060 at a wavelength of $\lambda=762$ nm. Wider air holes ($W_{x,z}=0.5W, 0.5a$) realized on a separate diamond sample (with higher surface roughness) resulted in an experimental quality factor of $Q$~2,100 in the heterostructure cavity (see Fig. 3c), and a $Q$~1,200 for the parabolic cavity (see Fig. 3d) at the NV ZPL wavelength ($\lambda=638$ nm) for $a=240$ nm. The measured resonant wavelengths show close correspondence to FDTD simulations of the ideal nanobeam cavity ($\theta=45°$). However, the measured average $Q$ factors (ranging from 700 to 3000 depending on the design and the mode) are significantly lower than the values predicted by simulations for the lowest cavity modes (see Fig. 3c). We associate these higher losses and wavelength shifts with the non-ideal triangular beam profile, the asymmetry in the mask alignment that impacts cavity geometry, and surface defects on the Type-IIa diamond (see island-like substrate defects in Fig. 2a-6). We speculate that these quality factors could be increased in higher-purity diamond with lower surface roughness (Ra < 5 nm). We note that despite similar $Qs$, higher order cavity modes are usually delocalized from the cavity center, thus having higher mode volumes and nodes where emitters should not be placed. This may be an important consideration in NV-cavity interaction.

To create large-scale circuits while maintaining structural stability, we implemented a tapered support bridge structure[13] to connect triangular nanobeam waveguides and cavities. The support structure expands from the waveguide width $W=500$ nm to a full width $W_b$ quasi-adiabatically over a distance of $2$ μm ($L_B/2$) to a support of width $W_s$ to decrease the waveguide mode overlap with the support (see Fig. 4a). The parameter space of the support geometry was constrained because we found that for fabricated structures with $L_b=8$ μm, $W_b = 1.3$ μm and $W_s=200-300$ nm over 85% of the beams were broken after mask removal in KOH, while decreasing $L_b$ to $4$ μm resulted in more stable structures with yield over 95%. Furthermore, while decreasing $W_s$ towards 100nm decreases the perturbation of the waveguide mode by the support (and thus the loss), it also makes the masks more fragile and reduces device yield. Thus, simulations for the optimized design shown in Fig. 4b examine both $W_s=100$ nm and $W_s=200$ nm, assuming $\theta=45°$.

We also introduce a design with supports on only one side of the nanobeam waveguide (Fig. 4a), which produces less perturbation on the waveguide mode. For each pair of parameters ($W_s$ and 1- or 2-sided support), the coefficients of the quintic polynomial representing the taper for maximum transmission in 3D FDTD are optimized, while constraining the derivative of the width at the beginning and end of the taper to be zero. Fig. 4b displays these maxima among a range of other $W_b$. We find that with a one-sided support with $W_s = 200$ nm, we can achieve a transmission of 97.4%, or -0.12 dB, while decreasing $W_s$ to $100$ nm results in transmission of 98.78%, or -0.05 dB. Beginning with the optimized designs based on these four geometries, *$\theta$ is varied* in simulation to determine its effect on the transmission efficiency. The results are shown in table in Fig. 4b. This transmission data show that the optimal taper shape is not necessarily similar across different etch angles. The taper parameters must be reoptimized for $\theta$ varying considerably from 45°. Still, even without optimization, transmission over 85% is achieved. Fig. 4c shows the fabricated support structure for $W_s=200$ nm. This structure is mechanically stable, thus enabling the fabrication of large networks.

In summary, we have shown the design of integrated waveguide-cavity networks and have fabricated the essential network elements – cavities, waveguides, and waveguide suspension structures, in CVD diamond using a mechanically transferred silicon hard-mask process that requires fewer processing steps on the diamond than previously demonstrated process[10] and enables etch selectivity[13] in excess of 99:1.  The experimentally measured $Q$ factors are in excess of 3000. The implementation of extended photonic networks in single crystal diamond is a promising platform for the realization of many individually controllable, entangled quantum memories.

After submitting this work, a preprint of a different approach towards realization of triangular nanobeams has been brought to our attention[14].

This work was supported in part by the Air Force Office of Scientific Research (PECASE Grant No. FA9550-11-1-0014, supervised by Gernot Pomrenke). This Research was carried out in part at the Center for Functional Nanomaterials, Brookhaven National Laboratory, which is supported by the U.S. Department of Energy, Office of Basic Energy Sciences, under Contract No. DE-AC02-98CH10886. The Technion contribution was supported in part by the German Israel Foundation (GIF) Contract No. I-1026-9.14/2009. T.S. was supported by the Alexander von Humboldt Foundation. E.C acknowledges support from the NASA Office of the Chief Technologist's Space Technology Research Fellowship. S. M. was supported in part by the AFOSR Quantum Memories MURI and the NSF Interdisciplinary Quantum Information Science and Engineering IGERT. The authors would like to thank Mircea Cotlet and Michael Walsh for their help and valuable discussions.

**Figure Captions:**

FIG. 1. (a) Schematics of the triangular nanobeam geometry. (b) Parabolic cavity formation. (c) Cavity parameter optimization for $Q$ and $V_m$ vs. $W_x/W$ for $W=3.82a$ $H=1.1a$, $W=3.82a$, $L=60a$, $W_z=0.5a$ (top); $Q$, $V_m$ vs. $W$ for $W_x=0.525W$ (center); $Q$, $V_m$ vs. $H$ for $W_x=0.525W$, $W=2a$ (bottom). (d) Normalized electric field density of the optimal mode at $z=0$ and $y=0.55a$.

FIG. 2. (a) The principal nanobeam fabrication process with optical and electron micrographs. (b) High magnification zoom-in of broken beam walls demonstrating low wall roughness. (c) Focused-Ion-Beam (FIB) made cross-section of the beam covered with platinum and diamond re-deposition.

FIG. 3. (a) SEM zoomed-in images of characteristic cavities with $W_z=0.3a$, $0.5a$ (left) and $W_z=0.5a$ mask (right). (b) Spectra taken from a parabolic cavity ($a=240$ nm, $W_z=0.3a$) (top). The Lorentzian fit yields $Q=3,057$ at $\lambda=761.96$ nm (bottom). (c) Spectrum of a double heterostructure cavity spectra ($a=240$ nm, $W_z=0.5a$). Peaks *iii* to *vi* are cavity resonances (modes *4-1*, respectively) with their measured and simulated values, i is the diamond Raman line (532 nm excitation), and *ii* and *vii* is background from fluorescence. (d) $\lambda$ and $Q$ dependence on the lattice constant $a$ for the first four modes in parabolic (left) and linear (right) defects (The data are averaged over an ensemble of 14 parabolic and 10 heterostructure cavities with identical designs).

FIG. 4. (a) Design of the modified one- and two- sided (semi-transparent) supporting interconnect (b) Transmission efficiency of the optimized structure in (a) for fundamental waveguide modes as a function of $W_b$ with $W_s=100, 200$ nm for etch angle $\theta =45°$ (top). Transmission efficiencies summary after modifying $\theta$ of each of the four optimized designs (bottom) (c) SEM of supports structures with $W_s=200$ nm.

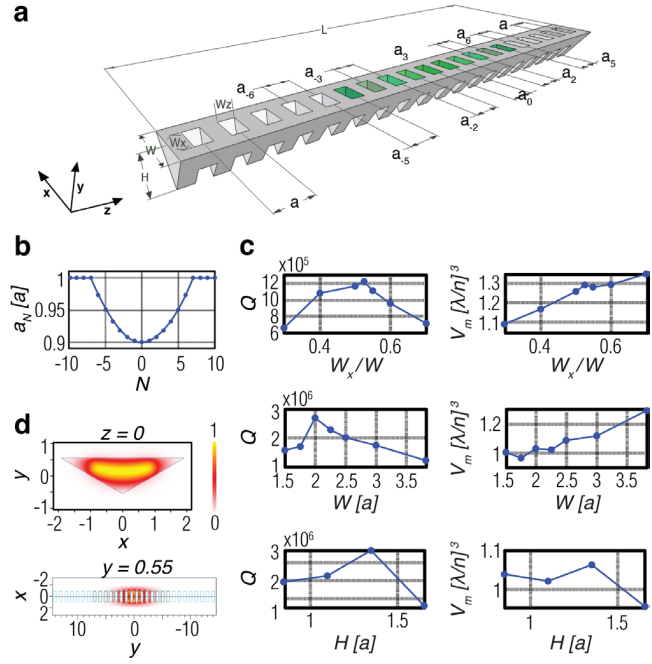

FIG. 1. (a) Schematics of the triangular nanobeam geometry. (b) Parabolic cavity formation. (c) Cavity parameter optimization for $Q$ and $V_m$ vs. $W_x/W$ for $W=3.82a$ $H=1.1a$, $W=3.82a$, $L=60a$, $W_z=0.5a$ (top); $Q$, $V_m$ vs. $W$ for $W_x=0.525W$ (center); $Q$, $V_m$ vs. $H$ for $W_x=0.525W$, $W=2a$ (bottom). (d) Normalized electric field density of the optimal mode at $z=0$ and $y=0.55a$.

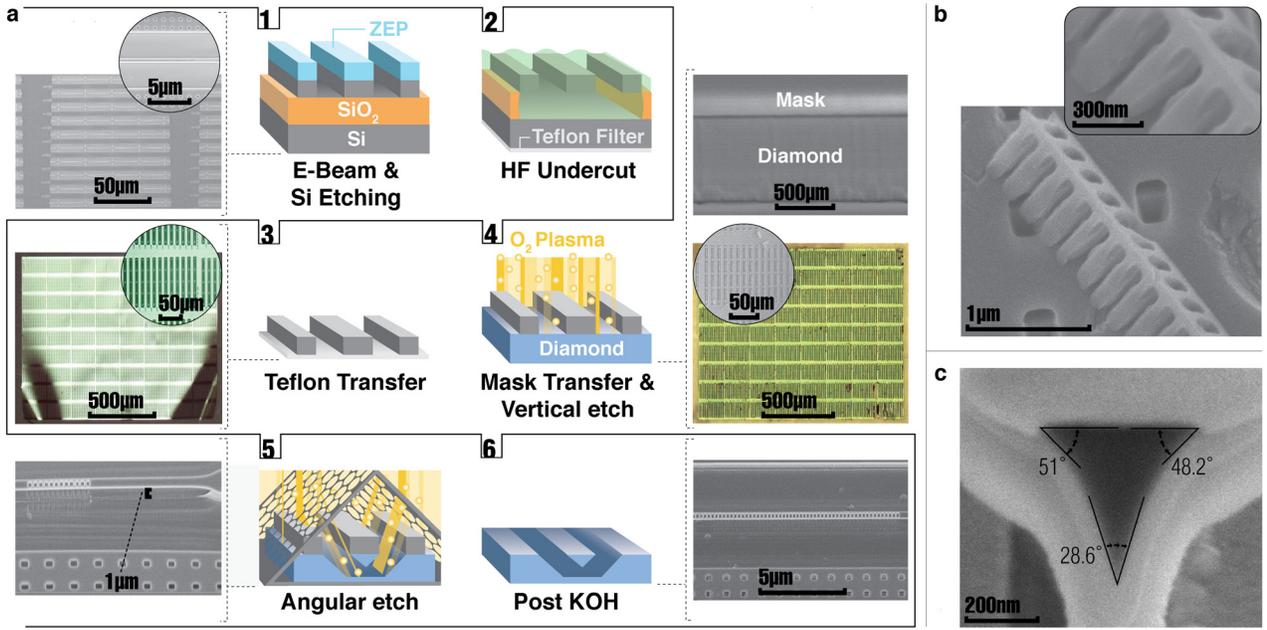

FIG. 2. (a) The principal nanobeam fabrication process with optical and electron micrographs. (b) High magnification zoom-in of broken beam walls demonstrating low wall roughness. (c) Focused-Ion-Beam (FIB) made cross-section of the beam covered with platinum and diamond re-deposition.

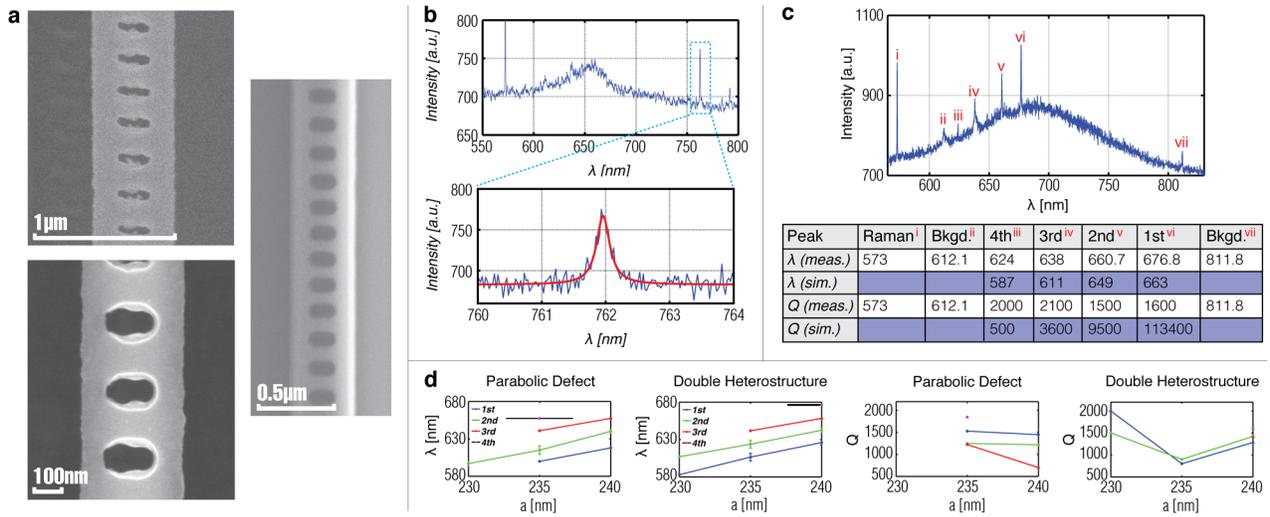

FIG. 3. (a) SEM zoomed-in images of characteristic cavities with $W_z=0.3a, 0.5a$ (left) and $W_z=0.5a$ mask (right). (b) Spectra taken from a parabolic cavity ($a=240$ nm, $W_z=0.3a$) (top). The Lorentzian fit yields $Q=3,057$ at $\lambda=761.96$ nm (bottom). (c) Spectrum of a double heterostructure cavity spectra ($a=240$ nm, $W_z=0.5a$). Peaks *iii* to *vi* are cavity resonances (modes *4-1*, respectively) with their measured and simulated values, *i* is the diamond Raman line (532 nm excitation), and *ii* and *vii* is background from fluorescence. (d) $\lambda$ and $Q$ dependence on the lattice constant $a$ for the first four modes in parabolic (left) and linear (right) defects (The data are averaged over an ensemble of 14 parabolic and 10 heterostructure cavities with identical designs).

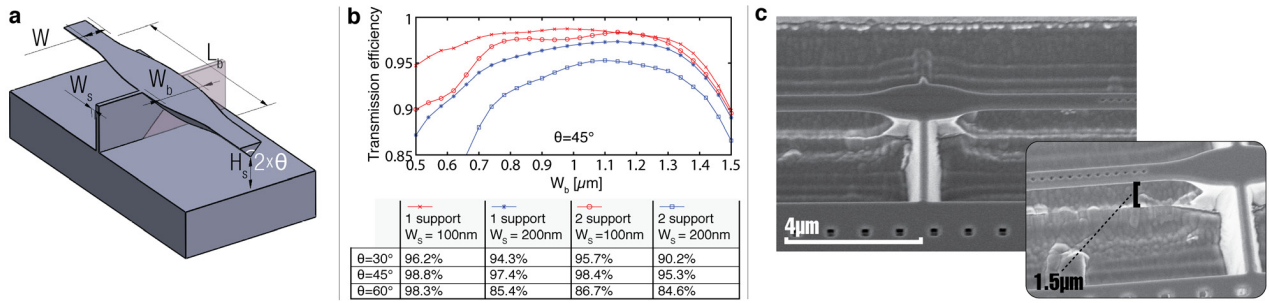

FIG. 4. (a) Design of the modified one- and two- sided (semi-transparent) supporting interconnect (b) Transmission efficiency of the optimized structure in (a) for fundamental waveguide modes as a function of $W_b$ with $W_s=100, 200$ nm for etch angle $\theta=45°$ (top). Transmission efficiencies summary after modifying $\theta$ of each of the four optimized designs (bottom) (c) SEM of supports structures with $W_s=200$ nm.